\documentclass[aps,prx,amstext,amsmath,twocolumn,superscriptaddress,showpacs]{revtex4-1}
\usepackage{graphicx}
\usepackage{color}

\newcommand{\figref}[1]{Fig.~\ref{#1}}
\newcommand{\Figref}[1]{Figure ~\ref{#1}}

\begin{document}

\title{Spatially - resolved study of the Meissner effect in superconductors using NV-centers-in-diamond optical magnetometry}

\author{N.~M.~Nusran}
\affiliation{Ames Laboratory, Ames, IA 50011}
\affiliation{Department of Physics $\&$ Astronomy, Iowa State University, Ames, IA 50011}

\author{K.~R.~Joshi}
\affiliation{Ames Laboratory, Ames, IA 50011}
\affiliation{Department of Physics $\&$ Astronomy, Iowa State University, Ames, IA 50011}

\author{K.~Cho}
\affiliation{Ames Laboratory, Ames, IA 50011}
\affiliation{Department of Physics $\&$ Astronomy, Iowa State University, Ames, IA 50011}

\author{M.~A.~Tanatar}
%\email{tanatar@ameslab.gov}
\affiliation{Ames Laboratory, Ames, IA 50011}
\affiliation{Department of Physics $\&$ Astronomy, Iowa State University, Ames, IA 50011}

\author{W.~R.~Meier}
\affiliation{Ames Laboratory, Ames, IA 50011}
\affiliation{Department of Physics $\&$ Astronomy, Iowa State University, Ames, IA 50011}

\author{S.~L.~Bud'ko}
\affiliation{Ames Laboratory, Ames, IA 50011}
\affiliation{Department of Physics $\&$ Astronomy, Iowa State University, Ames, IA 50011}

\author{P.~C.~Canfield}
\affiliation{Ames Laboratory, Ames, IA 50011}
\affiliation{Department of Physics $\&$ Astronomy, Iowa State University, Ames, IA 50011}

\author{Y.~Liu}
%\email{yliu@ameslab.gov}
\affiliation{Ames Laboratory, Ames, IA 50011}

\author{T.~A.~Lograsso}
%\email{lograsso@ameslab.gov}
\affiliation{Ames Laboratory, Ames, IA 50011}

\author{R.~Prozorov}
\email[Corresponding author: ]{prozorov@ameslab.gov}
\affiliation{Ames Laboratory, Ames, IA 50011}
\affiliation{Department of Physics $\&$ Astronomy, Iowa State University, Ames, IA 50011}

\date{\today}

\begin{abstract}
Non-invasive magnetic field sensing using optically - detected magnetic resonance of nitrogen-vacancy (NV) centers in diamond was used to study spatial distribution of the magnetic induction upon penetration and expulsion of weak magnetic fields in several representative superconductors. Vector magnetic fields were measured on the surface of conventional, Pb and Nb, and unconventional, LuNi$_2$B$_2$C, Ba$_{0.6}$K$_{0.4}$Fe$_2$As$_2$, Ba(Fe$_{0.93}$Co$_{0.07}$)$_2$As$_2$, and CaKFe$_4$As$_4$, superconductors, with diffraction - limited spatial resolution using variable - temperature confocal system. Magnetic induction profiles across the crystal edges were measured in zero-field-cooled (ZFC) and field-cooled (FC) conditions. While all superconductors show nearly perfect screening of magnetic fields applied after cooling to temperatures well below the superconducting transition, $T_c$, a range of very different behaviors was observed for Meissner expulsion upon cooling in static magnetic field from above $T_c$. Substantial conventional Meissner expulsion is found in LuNi$_2$B$_2$C, paramagnetic Meissner effect (PME) is found in Nb, and virtually no expulsion  is observed in iron-based superconductors. In all cases, good correlation with macroscopic measurements of total magnetic moment is found.  Our measurements of the spatial distribution of magnetic induction provide insight into microscopic physics of the Meissner effect.

\end{abstract}

\maketitle

\section{Introduction}
\subsection{Meissner effect in superconductors}
\begin{figure}[tb]
\includegraphics[width=8.5cm]{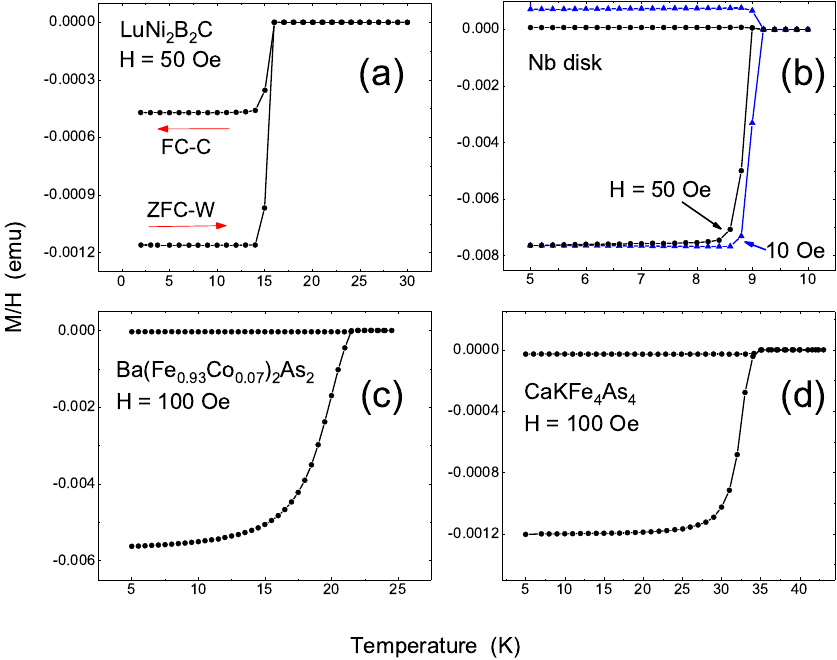}
	\caption{Temperature - dependent total magnetic moment measured using \emph{Quantum Design} MPMS. Shown are zero-field-cooling warming (ZFC-W) curves and field-cooling cooling (FC-C, Meissner expulsion) curves measured in single crystals of (a) LuNi$_{2}$B$_{2}$C borocarbide; (b) niobium; (c) Ba(Fe$_{0.93}$Co$_{0.07}$)$_2$As$_2$ and (d) CaKFe$_4$As$_4$ in magnetic fields used in the experiments below.}
	\label{MPMS}
\end{figure}

Contrary to simplified introductions into the subject of superconductivity, the expulsion of weak magnetic fields from a superconductor, known as the ``Meissner-Ochsenfeld Effect" or more often just as ``Meissner Effect (ME)", is still not fully explored both experimentally and theoretically when real samples of finite size and non-ellipsoidal shape are used. It is well established that weak magnetic field penetrates a homogeneous superconducting sample only to a small depth at the edges (London penetration depth and some vortices with density gradient proportional to pinning strength), setting the quantitative measure of a total diamagnetic moment corresponding to a complete flux expulsion. The distinct characteristic property of a superconductor, the Meissner effect, however, is the flux \emph{expulsion} upon cooling through the superconducting transition, $T_c$, in a magnetic field. In this case, measurements of the total magnetic moment range from a (very rare) \emph{complete} flux expulsion in clean type-I superconductors of ellipsoidal shape \cite{Prozorov05}, to nearly complete expulsion in pinning - free conventional type-II superconductors \cite{Huebener01,Tinkham96}, to practically no expulsion in iron pnictides \cite{Prozorov10}, to \emph{paramagnetic} Meissner effect observed in various materials with extreme sensitivity to disorder. This variety of behavior is shown in \figref{MPMS} where total magnetic moment of bulk superconducting samples was measured using \emph{Quantum Design} MPMS. Similar data for type-I Pb superconductor showing a complete Meissner expulsion are published in Ref. \onlinecite{Prozorov08,Prozorov07}. All samples were chemically homogeneous well - characterized single crystals and partial ``superconducting volume fraction'' due to poor quality or granularity can be excluded.
Note: we use the following terminology and abbreviations throughout the paper: (1) ZFC-W - sample is cooled in zero field to low temperature, $T<T_c$ and a magnetic field is applied. Then the measurements are taken on warming through $T_c$; (2) FC-C - the measurements are taken while the sample is cooled in a magnetic field; (3) ZFC - sample is cooled in zero field down to low temperature (4.2~K) and a magnetic field is applied. Then the measurements are taken at 4.2~K; (4) FC - sample is cooled in a magnetic field. Measurements are performed at 4.2~K.

Clearly, it is impossible to understand FC-C results without knowing the distribution of the magnetic induction throughout the sample, which had to be assumed in theoretical models. Thus spatially - resolved measurements are needed. There are a few reports of direct visualization of FC-C process using magneto-optical technique.\cite{Prozorov07,Prozorov05,Prozorov93} In most cases, however, spatially resolving probes have at least one limitation that hinders a full study of the FC state. Examples are limited sensitivity, insufficient spacial resolution, limited mapping area or invasive nature. Furthermore, in a usual experiment, the sample is a flat slab with large demagnetization effects that make field distribution highly non-uniform. Therefore, in the ideal case the measurement should be non-invasive and performed on a well - characterized sample with well - defined sharp edges (for example, as seen by electron microscopy), provide sufficient spacial resolution, be sensitive enough to detect magnetic field from a few vortices and the sample has to be stationary, because motion in a magnetic field may lead to effective field changes due to omnipresent in magnets field gradients.

In this report we use a novel non-invasive optical magnetometer that satisfies these restrictions to probe the structure of the Meissner effect in several superconductors and show how different the behavior is. In some cases, such as PME in Nb, we confirm the theory suggested by Koshelev and Larking that flux compression is the most likely scenario for the observed apparent paramagnetism \cite{Koshelev95}. In other, such as iron based superconductors, we simply confirm that Meissner expulsion is virtually absent on the scale of the whole sample \cite{Prozorov10}, but the reason is still unclear. And, indeed, we observe conventional Meissner expulsion in cases where it is expected, such as low pinning borocarbides.

\subsection{Optical magnetometer based on NV - centers in diamond}
\begin{figure*}[tb]
\includegraphics[width=14cm]{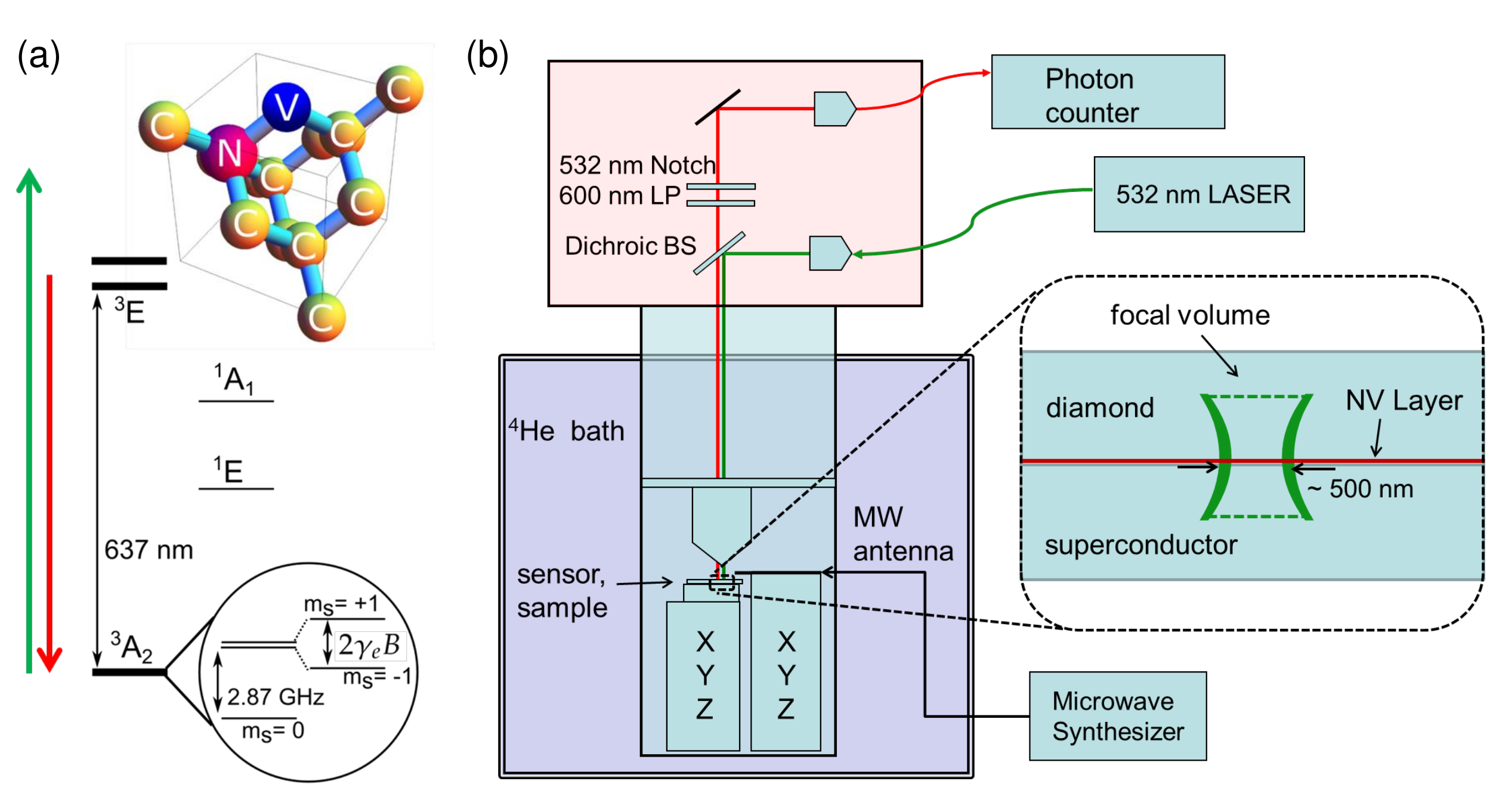}
	\caption{(a) Crystal structure of diamond with NV center and its energy levels structure. Under off-resonant 532~nm laser excitation, the NV emits red photons. Due to the additional relaxation path via the meta-stable states the $m_S=\pm1$ levels fluorescence less than $m_S=0$ spin sub level. (b) Experimental setup: A low temperature confocal fluorescence microscope optimized for NV detection. The spatial resolution of the diamond sensor is determined by the convolution of the focal volume with the NV distribution in the diamond plate, leading to a disk-like probe of thickness $\approx 20$~nm and diameter $\approx 500$~nm. See Methods for more details.}
\label{nv}
\end{figure*}

The nitrogen-vacancy (NV) center is a point defect in the diamond lattice that consists of a nearest neighbor pair of a substitutional nitrogen atom and a lattice vacancy shown schematically in \figref{nv}(a). With an additional acquired electron, NV center has a spin triplet, $S=1$, ground state. (In this paper, we exclusively consider the negatively charged NV centers and simply refer to it as the NV center). When excited to a higher energy level (i.e., by a 532~nm green laser) from the $m_S=0$ spin projection ground state, the relaxation back to $m_S=0$ proceeds through spin - conserving cyclic transitions emitting red photons. However, if excited from $m_S=\pm$1 levels NV center can also relax via the meta-stable (dark) states to $m_S=0$ resulting a reduced red fluorescence rate. This spin - dependent fluorescence allows for optical detection of the magnetic spin resonance (ODMR) by sweeping frequency of microwave radiation. When the frequency matches the energy difference between $m_S=$0 and $m_S=\pm$1 levels, i.e., when electron spin resonance (ESR) occurs, the fluorescence rate is minimal. In the presence of magnetic field, the frequency of ESR signal is shifted owing to the Zeeman effect, thus the change of resonance frequency can be used as a probe to accurately measure the local magnetic field. As a consequence of long coherence time, convenient energy levels spacing and several important advances in the measurement protocols and sequences, NV - centers in diamond are now emerging as a very promising candidate for \emph{non-invasive} optical magnetometry with nano-scale spatial resolution.\cite{Pelliccione16,Theil16,Grinolds14,Grinolds13,Rondin13,Rondin12} A detailed review of the NV-centers and NV magnetometry can be found in Refs.\cite{Doherty13, Rondin14}. The non-invasive nature of the technique is very important for probing delicate states, especially those where quantum coherence is important, where conventional measurements may alter the state of the studied system.

The effective ground state Hamiltonian of an NV - center is given by:

\begin{equation}
H_{eff} = D S_z^2 +E(S_x^2-S_y^2)- \gamma_e \vec{S} \cdot \vec{B}
\end{equation}

\noindent where $\gamma_e \approx 28$~GHz/T is the gyro-magnetic ratio of the NV electronic spin, $D =2.87$~GHz and $E \approx 5$~MHz are axial and off-axial zero-field splitting parameters respectively.\cite{Rondin14} In this work, we measure the magnetic fields through the detection of Zeeman splitting observed in the ODMR spectra.

The experimental apparatus incorporates a confocal microscope optimized for NV fluorescence detection. The fluorescence is stimulated by the green off-resonant 532~nm laser excitation and low - energy levels are populated by the microwave radiation applied using a single silver wire loop antenna coupled to a MW frequency generator. A thin diamond plate with an ensemble of NV-centers embedded near the surface ($\sim$ 20~nm depth) is used as the magneto - optical sensor. The spatial resolution of the sensor is determined by the effective size of the probe, which is essentially a convolution of the focal volume with the NV distribution in the diamond plate. This leads to a disk-like probing volume of thickness $\approx 20$~nm and diameter $\approx 500$~nm. See Methods for more details.

\begin{figure}[h!]
\includegraphics[width=8cm]{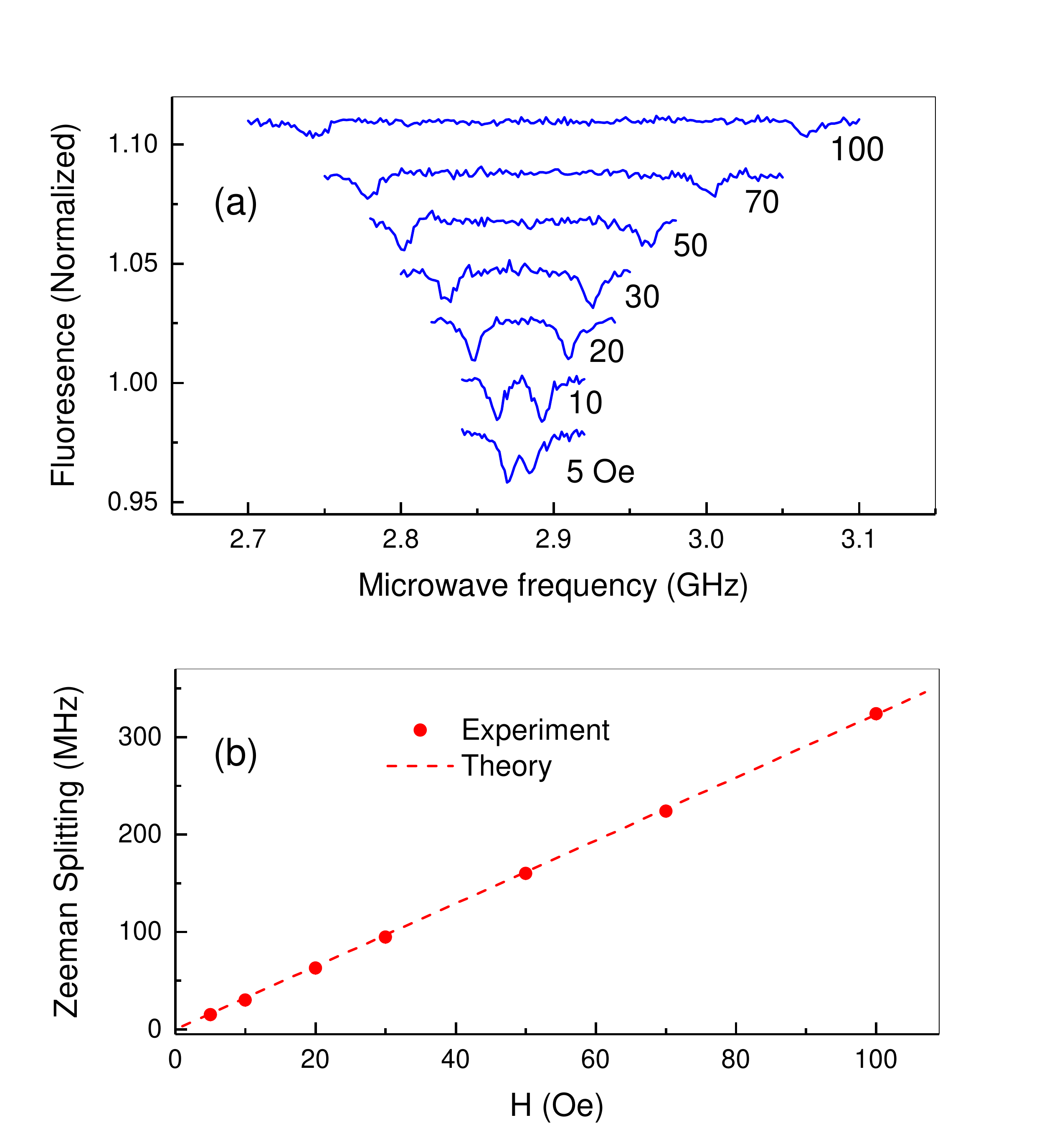}
	\caption{(a) ODMR spectrum of the sensor without a sample for several values of the external magnetic field. For a single crystalline diamond plate with [100] surface placed normal to the field, all four NV orientations result in the same Zeeman splitting. (b) Experimentally measured Zeeman splitting vs. the applied magnetic field. The dashed line is the theoretically anticipated curve, $3.233$~MHz/Oe. (See Appendix)}
	\label{calib}
\end{figure}

\Figref{calib}(a), shows the ODMR spectrum for different values of the external magnetic magnetic field without the sample. Only one pair of splitting in the ODMR spectrum is observed, because in the case of a single crystalline diamond plate with [100] surface placed normal to the field, $\vec{B}=(0,0,B_z)$, all the four possible NV orientations result in the same Zeeman splitting. See Appendix for more details. The slight broadening of the ODMR resonances at higher magnetic fields is most likely due to a small mismatch in the normal direction of the crystalline plane with respect to the magnetic field. Moreover, the NV centers here always experience a perpendicular component of the magnetic field with respect to their N-V axis. This leads to level mixing at higher magnetic fields, resulting in lower contrast of the ODMR signal\cite{Epstein05,Lai09} limiting the practically measurable fields to about 200~Oe. The sensor also looses its sensitivity when Zeeman splitting is comparable with the off-axis splitting parameter $E$, leading to a lower-bound of approximately $2$~Oe for the directly measurable fields.\cite{Rondin14}

\section{Results and discussion}

\subsection{LuNi$_2$B$_2$C}

\begin{figure}[tb]
\includegraphics[width=8cm]{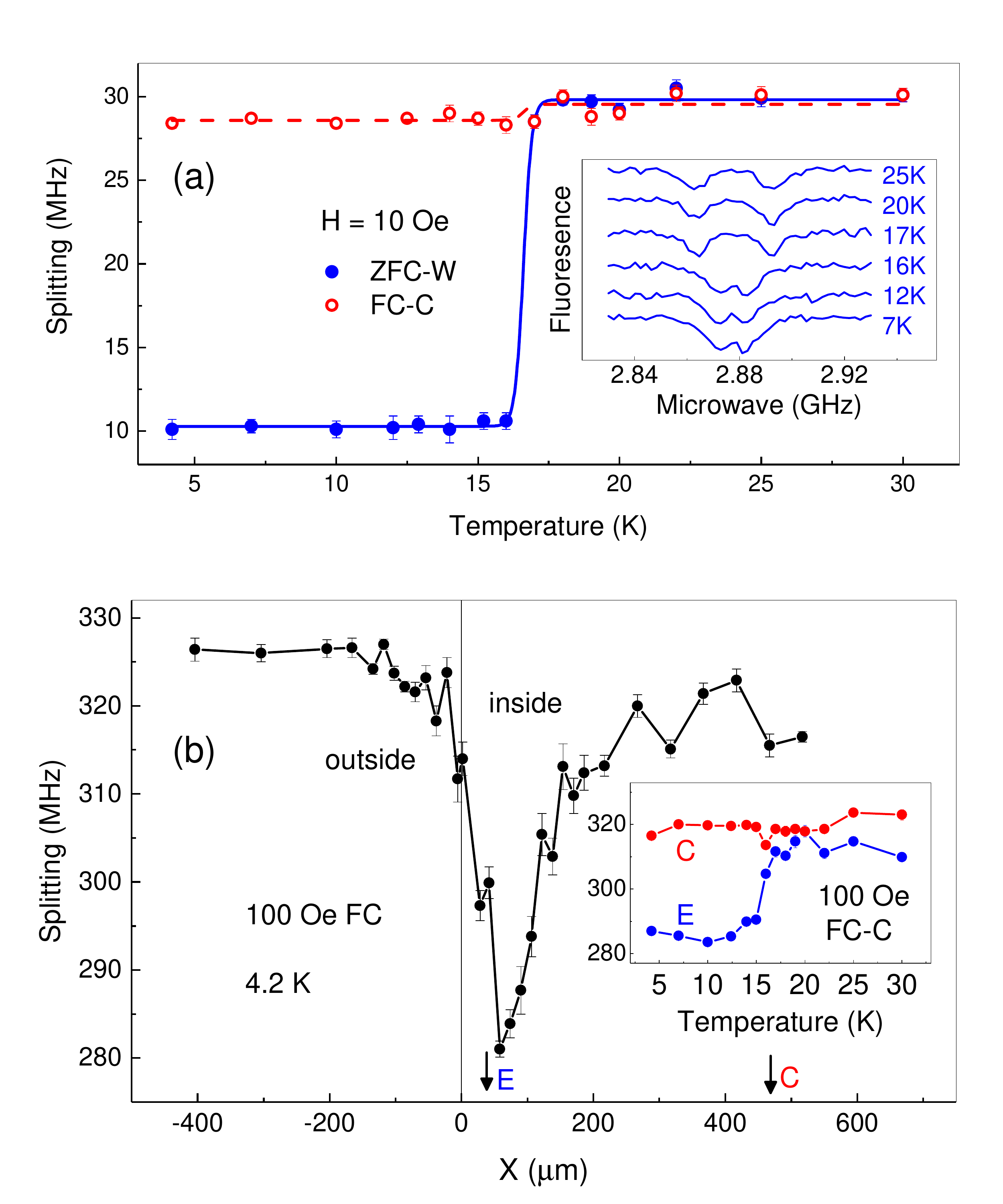}
	\caption{(a) Superconducting transition in LuNi$_2$B$_2$C measured at the sample center. Blue filled circles show Zeeman splitting obtained from ODMR (inset) with increasing temperature. Blue solid line is a best fit to a sigmoid function with  $T_c~=~16.6 \pm 0.1$~K. The open red circles and dashed line show the data and the fit for the FC-C measurements. (b) FC profile at 100~Oe across LuNi$_2$B$_2$C at 4.2~K. Sample is located at $x \geq 0$. Meissner expulsion is clearly visible near the edge of the sample, $x=$0. (Inset) Zeeman splitting with decreasing temperature at two different points: near the edge (E) and near the center (C) of the superconductor.}
	\label{BC}
\end{figure}

We begin with the single crystal LuNi$_2$B$_2$C - a type-II superconductor with low vortex pinning strength.\cite{Canfield97,Cava94} Filled blue symbols in the main panel of \figref{BC} show ODMR measured on warming after a small magnetic field of 10~Oe was applied at 4.2~K to which the sample was cooled without field (ZFC). Open red circles show the measurements performed on cooling the sample from above $T_c$ in the same field of 10~Oe. To avoid complications related to demagnetization, ODMR was measured near the sample center. Blue solid curve shows the fitting for the ZFC data to a sigmoid function\cite{Waxman14}:

\begin{equation}
S(T) = \frac{a}{1+\exp{[\frac{-(T-T_c)}{\delta T_c}]}} + b
\label{sigmoid}
\end{equation}

\noindent where $a,~b,~T_c,$ and $\delta T_c$ are fitting parameters, we obtain the critical temperature $T_c~=~16.6 \pm 0.1$~K, which is consistent with the literature.\cite{Canfield97}

Magnetic induction profile across the sample after cooling in a 100 Oe magnetic field from above $T_c$ to 4.2~K is shown in \figref{BC}(b). This profoundly non-monotonic spatial distribution was previously observed in magneto-optical experiments with bithmuth - doped iron garnet indicators in clean Y-Ba-Cu-O crystals.\cite{Prozorov92} This dome-like shape is induced by the competition between temperature - dependent critical current and temperature - dependent London penetration depth. As expected Meissner currents are significant only near the edges and the corresponding diamagnetic susceptibility, $4 \pi \chi = V^{-1}\int_{all space}(B-H)dV$ in the whole space (inside and outside the sample)  will be less than ideal.

\subsection{Single crystal Pb}

\begin{figure}[tb]
\includegraphics[width=8cm]{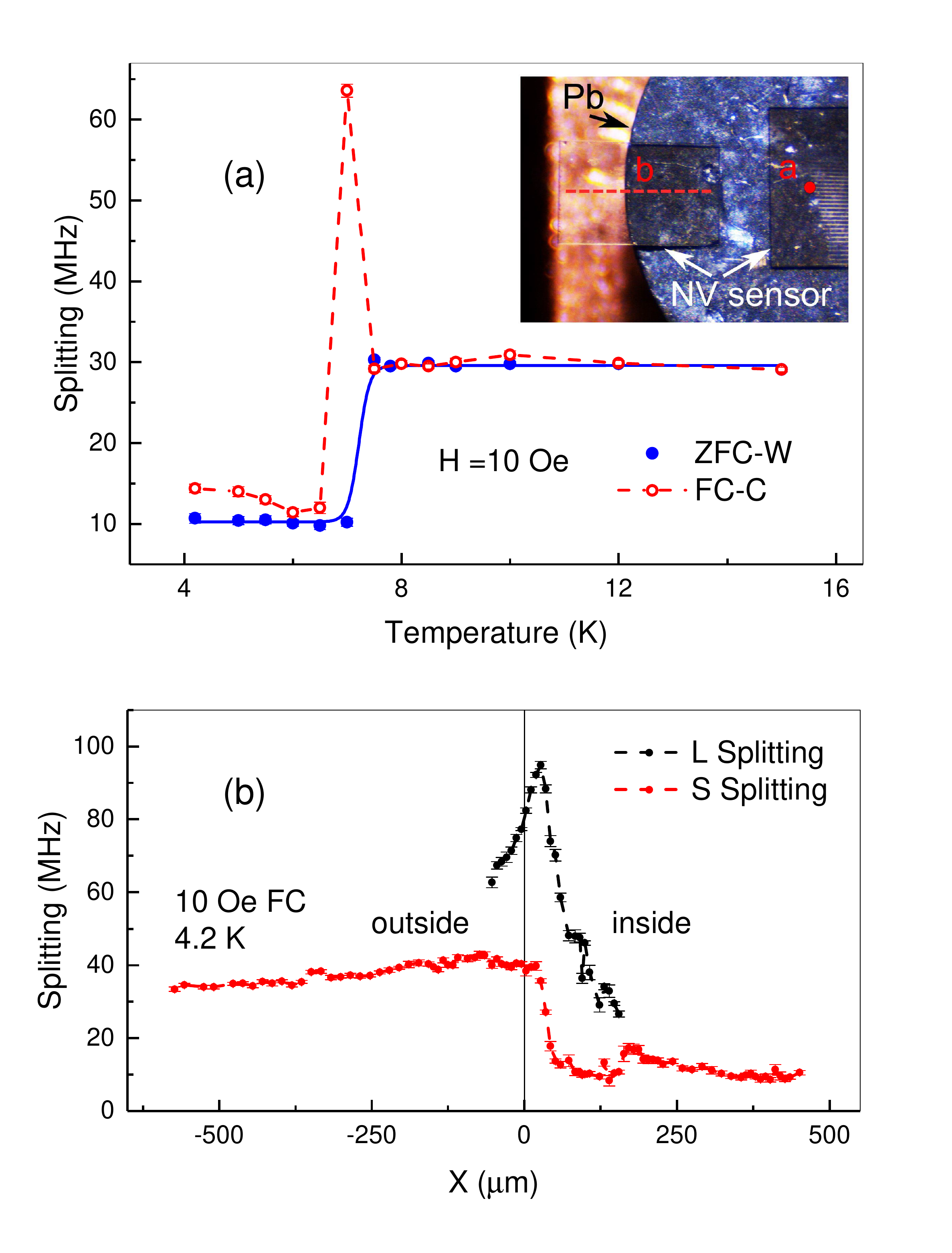}
	\caption{Panel (a) ZFC-W and FC-C measurements at point ``a'' shown in the inset. Panel (b) Magnetic flux profile at 10~Oe applied field measured along the path marked ``b'' in the inset.}
	\label{pb}
\end{figure}

For comparison with type-II superconductors, we studied Meissner expulsion in pure Pb crystal previously used in extensive magneto - optical studies and where a complete Meissner expulsion was observed both in measurements of total magnetic moment and directly with spatially resolved measurements, thanks to a large - scale laminar flux structure as opposed to sub-$\mu$m sized Abrikosov vortices Refs.\cite{Prozorov08,Prozorov07}. \Figref{pb} shows NV measurements on a disk - shaped Pb single crystal. The inset shows diamond slab placed on top of the sample. Panel (a) shows superconducting phase transition measured in a magnetic field of 10~Oe at the sample center (point ``a'' in the inset). The peak in FC-C curve just below $T_c$ is most likely due to the crossing of the measurement point by a normal phase lamella \cite{Prozorov07}. Panel (b) shows FC profile across the sample edge along the path ``b'' of the inset. As discussed in Appendix, two different sets of the resonance splittings are related to normal and longitudinal components of the magnetic induction at the location of NV centers. Meissner expulsion is clearly observed inside the sample and some variation is due to normal lamina pinned inside.

\newpage
\subsection{Single crystal Nb}

\begin{figure}[tb]
\includegraphics[width=9cm]{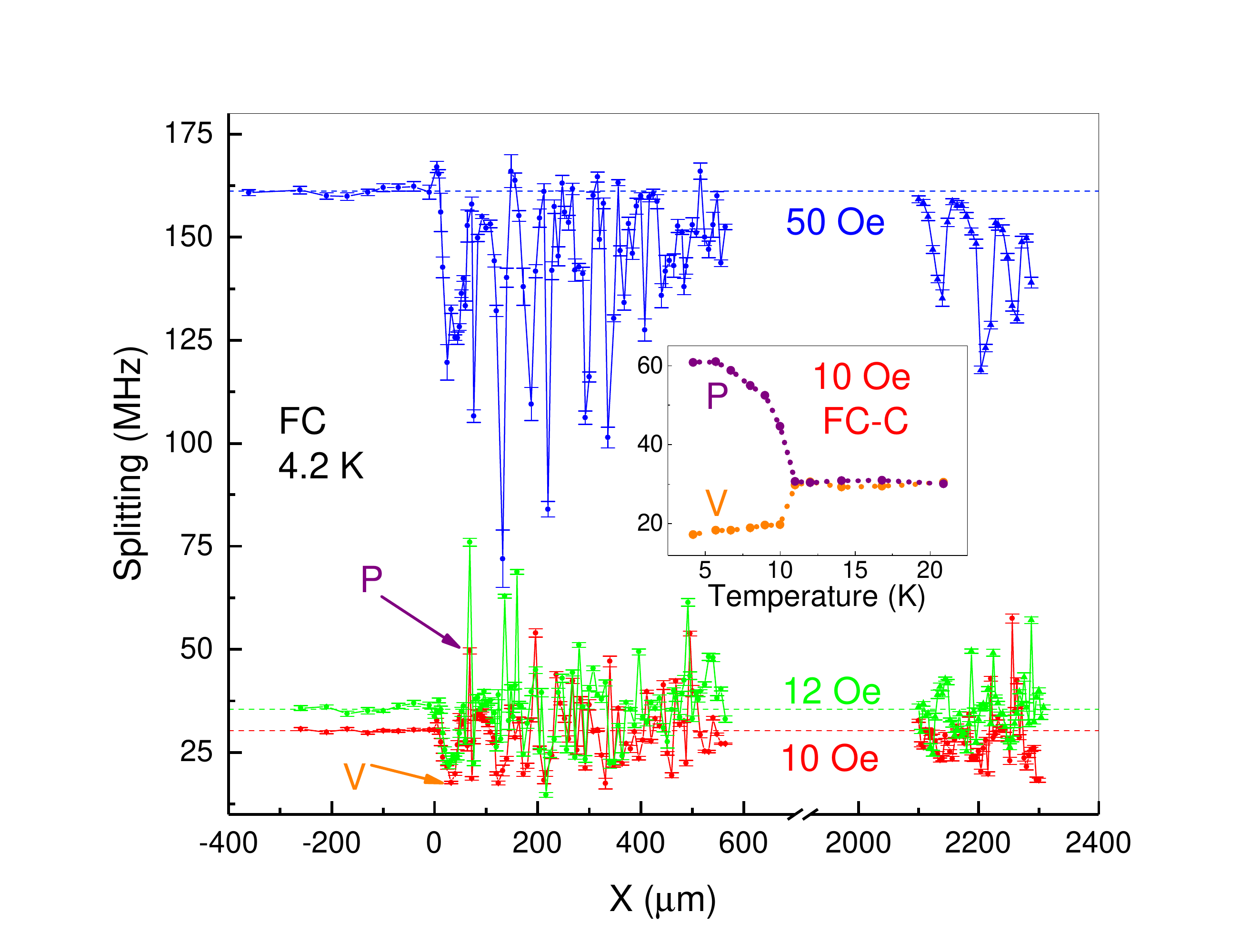}
	\caption{FC-C profiles of the magnetic induction for applied magnetic fields of 10, 12, and 50 Oe in Nb disk - shaped crystal measured at 4.2~K. Paramagnetic Meissner effect (PME) is observed in various random regions for lower magnetic fields. Cooling in 10 and 12 Oe fields results in very similar profiles. (inset) Temperature - resolved measurement in a 10 Oe magnetic field upon FC-C repeated at a peak (``P'') and a valley positions (``V'') shown by arrows in the main panel.}
	\label{nb}
\end{figure}

An obvious choice of a conventional type-II superconductor would be niobium. However, this material is far from being conventional as far as its properties in a magnetic field are concerned. In addition to a well documented so-called ``paramagnetic'' Meissner effect in low magnetic fields\cite{Kostic96} it also exhibits a huge increase of the upper critical field with disorder and shows collapsing in a catastrophic manner critical state.\cite{Prozorov06} \Figref{nb} shows field - cooled flux profiles across a 5~mm diameter and 1~mm thick Nb disk for three values of the applied magnetic field. The inset shows temperature - dependent signal measured at two locations, ``P'' and ``V'' corresponding to local maximum and local minimum as marked on the lower curve. Clearly, Nb does not exhibit a uniform Meissner expulsion upon field cooling, although at higher fields (50 Oe), the mean value is less than applied field implying negative total magnetic moment. For lower applied magnetic fields, \emph{an increase} of the local magnetic induction values is observed at many randomly appearing regions. The mean value of the induction is greater than the applied field and, therefore, total magnetic moment will be positive, - phenomenon known as the ``paramagnetic'' Meissner effect (PME).\cite{He96, Kostic96, Okram97, Kumar15} This is directly seen in our \figref{MPMS} (b). By repeating the experiment in the same and slightly different fields we observed that the structure of the magnetic induction modulation is not stochastic and is quite reproducible (see comparison of 10 and 12 Oe in \figref{nb}). This implies that the PME in Nb is due to some variation in $T_c$ and/or critical current density (as noted above, Nb shows very strong response to small amounts of disorder). Although a comprehensive study on PME is beyond the scope of this paper, we believe that inhomogeneous cooling and flux compression is the cause of PME observed in Nb as proposed theoretically by Koshelev, Larkin and Vinokur.\cite{Koshelev95, He96, Kostic96} For a detailed review on PME, we refer the reader to Ref.\cite{Li03}.

\subsection{Iron pnictides}
\subsubsection{Ba(Fe$_{0.93}$Co$_{0.07}$)$_2$As$_2$}
\begin{figure}[tb]
\includegraphics[width=8cm]{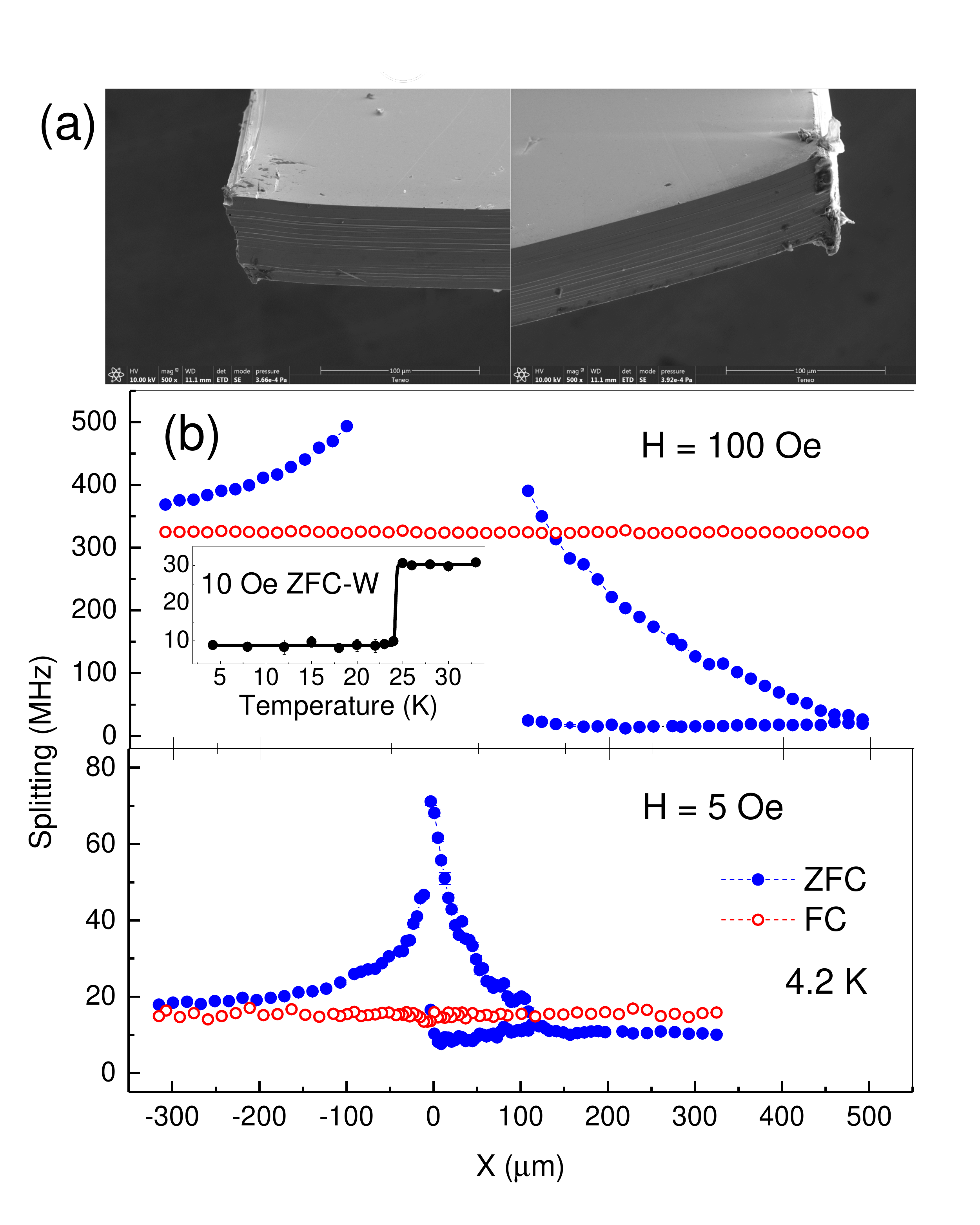}
	\caption{(a) Scanning electron microscope (SEM) images show the measured Ba(Fe$_{0.93}$Co$_{0.07}$)$_2$As$_2$ crystal with a reasonably rectangular shape and smooth surface. (b) FC and ZFC profiles for 100~Oe (top) and 5~Oe (bottom) measured at 4.2 K. $x < 0 (>0)$ is outside (inside) the sample. Inset shows  superconducting phase transition at  $T_c~\approx~24$~K.}
	\label{baco}
\end{figure}

Iron based superconductors are of immense interest for their various unusual properties. A particular interest here is anomalous Meissner effect identified by our group in these materials from the measurements of total magnetization.\cite{Prozorov10} As discussed in the introduction, such measurements always leave room for possible artifact and should be supported by the spatially - resolved techniques. Unfortunately, magneto - optical imaging is either not sensitive enough or there is truly no substantial Meissner expulsion in iron pnictides. First we study single crystal of Ba(Fe$_{0.93}$Co$_{0.07}$)$_2$As$_2$. Iron pnictides are known for their often very layered morphology and we examine the samples in a scanning electron microscope (SEM) and choose one with well-defined rectangular shape and good surface and edge (\figref{baco}(a)). The dimensions of the sample studied here are $\sim 1 \times 1.1 \times 0.06$~mm$^3$.

The inset in \figref{baco}(b) shows superconducting phase transition at $T_c~\approx~24$~K measured near the sample center, consistent with previous measurements of the same composition\cite{Gordon09}. \Figref{baco}(b) compares FC (solid red circles) and ZFC (filled blue circles) Zeeman splitting profiles at 100~Oe (top) and 5~Oe (bottom) applied magnetic fields. As expected, the two Zeeman pairs were observed in the ZFC profiles near the edge due to change in direction of the net magnetic induction field, caused by the shielding currents. Importantly, the FC profiles show no Meissner expulsion at all. This is consistent with the global measurements reported for this system\cite{Prozorov10}, but is very different from ordinary Meissner effect in LuNi$_2$B$_2$C or PME behavior in Nb shown above. There is no clear explanation for this effect, but we confirm its existence in spatially resolved measurements on a stationary sample.

\subsubsection{CaKFe$_4$As$_4$}

\begin{figure}[tb]
\includegraphics[width=9cm]{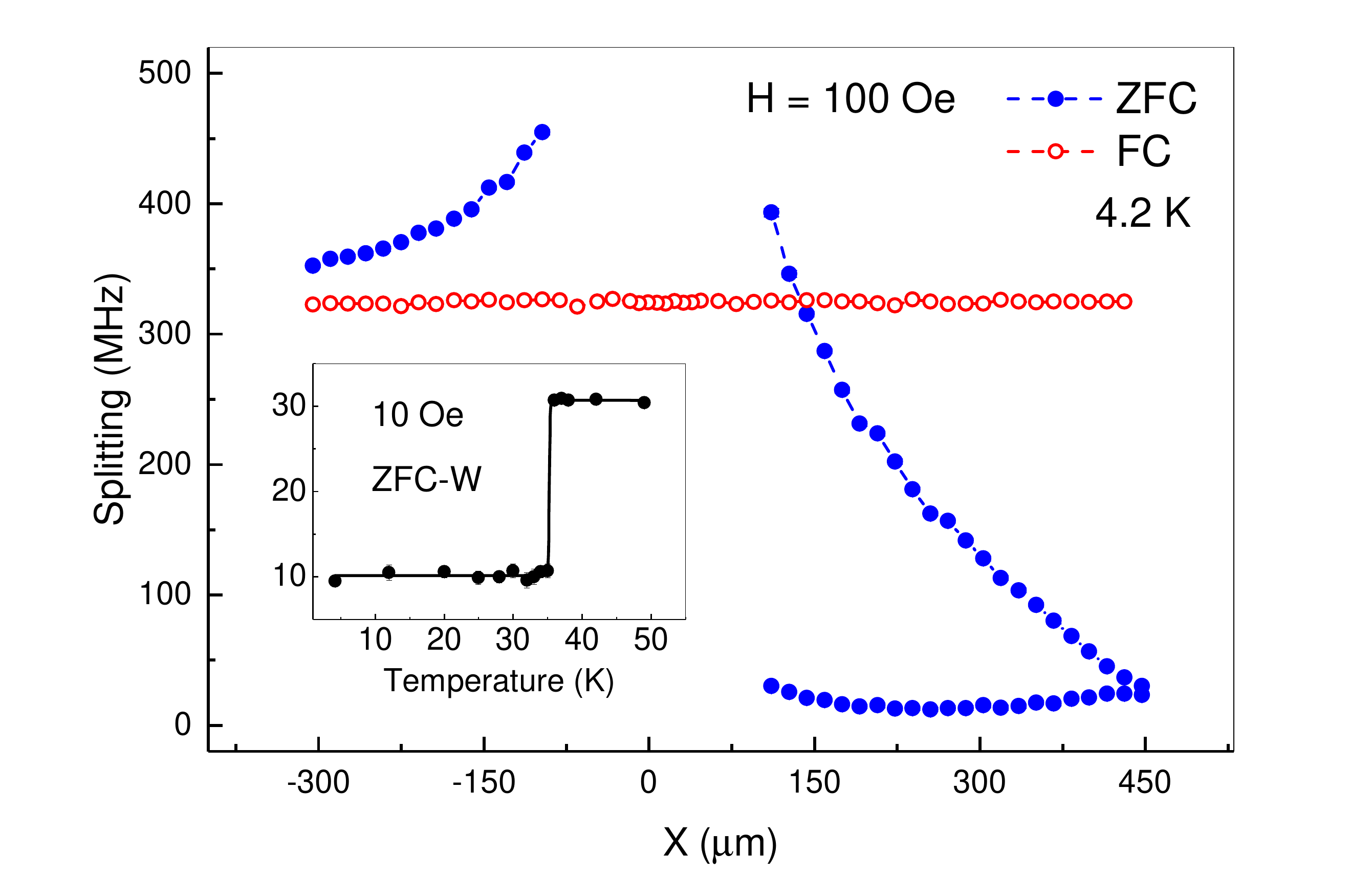}
	\caption{100~Oe FC and ZFC profiles at 4.2~K of CaKFe$_4$As$_4$. (inset) Detection of superconducting phase transition at $T_c~=~35.3 \pm 0.8$~K.}
	\label{cak1144}
\end{figure}

To gain further insight and see if the disorder from chemical substitution is to blame for the observed anomalous Meissner effect, we turned to a most recent addition to the pnictides family, stoichiometric CaKFe$_4$As$_4$.\cite{Meier16} The studied rectangular cross-section sample had dimensions of $ \sim 1 \times 1 \times 0.01$~mm$^3$. Inset in \figref{cak1144} shows the superconducting phase transition at the critical temperature $T_c~=~35.3 \pm 0.8$~K, consistent with previous measurements.\cite{Meier16} Despite being cleaner, the FC profile is flat and shows no change after cooling below $T_c$.

\subsubsection{Ba$_{0.6}$K$_{0.4}$Fe$_2$As$_2$}

\begin{figure}[tb]
\includegraphics[width=9cm]{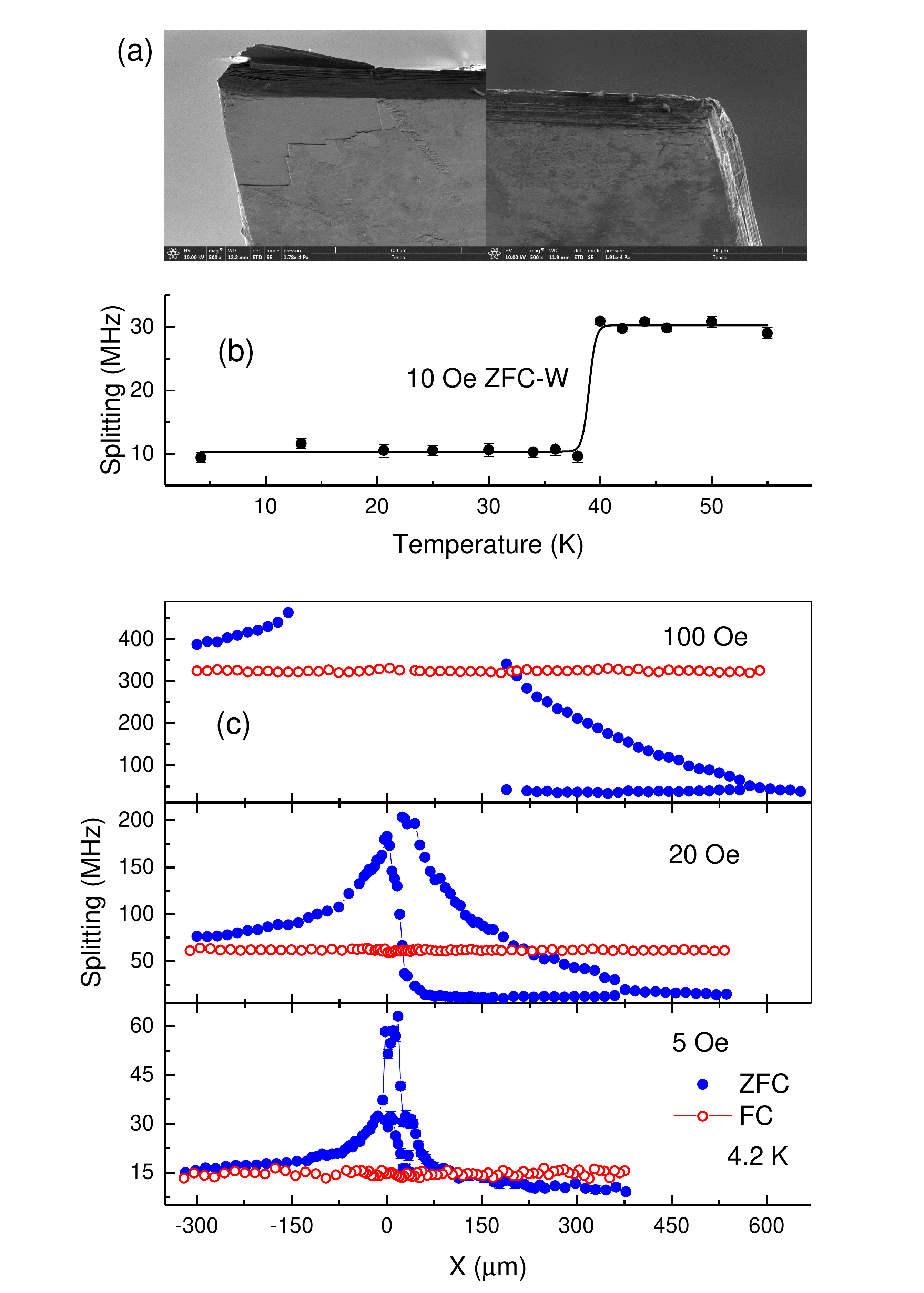}
	\caption{(a) SEM images of two sides of Ba$_{0.6}$K$_{0.4}$Fe$_2$As$_2$ crystal. The right side is chosen for profiling. (b) superconducting transition measured in the sample center in ZFC-W showing critical temperature $T_c~=~38.9 \pm 0.2$~K. (c) Comparison of ZFC and FC splitting profiles under 5, 20 and 100 Oe (from bottom to top) measured at 4.2~K.}
	\label{bak122}
\end{figure}

The study of ``122'' derived pnictide superconductors wouldn't be complete without hole - doped Ba$_{x}$K$_{1-x}$Fe$_2$As$_2$. Here we study the Meissner expulsion of  $ \sim 1.5 \times 1.4 \times 0.03$~mm$^3$ sized rectangular optimally - doped Ba$_{0.6}$K$_{0.4}$Fe$_2$As$_2$ sample. The SEM images are shown in \figref{bak122} and for the measurements we chosen the edge shown in the right panel.
\Figref{bak122}(b) shows superconducting phase transition at the critical temperature $T_c~=~38.9 \pm 0.2$~K, consistent with previous measurements.\cite{Cho16,Liu14} \Figref{bak122}(c) compares the FC and ZFC flux profiles at 5, 20 and 100~Oe applied magnetic fields at 4.2~K. As in the previous cases, the FC curves show neither Meissner expulsion nor PME behavior.

\section{Discussion}

Traditionally, the FC-C behavior in type II superconductors could be explained in a hand-waving argument as follows: when cooling from above $T_c$ in a magnetic field $H$, Abrikosov vortices are formed at $T_c(H)$ (or, equivalently, at $H=H_{c2}(T)$). In the ideal case without pinning Meissner expulsion is effective until the distance between vortices becomes of the order of London penetration depth, $\lambda$, because Meissner currents always present in so-called ``Meissner belt'' around the finite sample, also of width of the order of $\lambda$, push vortices \emph{into the sample}. Therefore, the degree of ultimate flux expulsion will always be less than 100\% and its value is determined by the complex competition between temperature dependent $\lambda(T)$ and $\xi(T)$ as well as demagnetization effects that renormalize magnetic field at the edges depending on the amount of the expelled flux. Adding temperature dependent pinning complicates things further. A detailed microscopic analysis of this situation is lacking and we hope that our measurements will provide motivation for such theoretical work. It is clear that a textbook statement that weak magnetic field is fully expelled from an ideal superconductor is only applicable for an infinite sample without demagnetization and boundaries. A finite specimen, even with zero pinning, will always have residual magnetic induction after field cooling of the order of the lower critical field, $H_{c1}$. The absence of Meissner expulsion at low fields and its appearance and increase at much higher fields (in a linear in field fashion, see Ref.\cite{Prozorov10}) implies that the degree of expulsion is scaled roughly as $H H_{c1}(0)/H_{c2}(0)$, so for iron pnictides it is very small, because $H_{c1}(0)/H_{c2}(0) \sim 10^{-3}-10^{-4}$. In the limit approaching type-I superconductors, $H_{c1}(0)/H_{c2}(0) \to 1$, a complete expulsion is expected and observed. Of course, there may be other factors affecting the behavior.

\section{Conclusion}

In conclusion, we developed low-temperature optical magnetometer based on ensembles of NV - centers in diamond crystal for studies of magnetic flux distribution in superconducting materials. We surveyed several superconducting systems that show both ordinary Meissner effect and anomalous paramagnetic Meissner effect in bulk magnetization measurements. Spatially - resolved information provides crucially important insight into the for the interpretation of the results.

\section*{Ackowledgements}
This work was supported by the U.S. Department of Energy (DOE), Office of Science, Basic Energy Sciences, Materials Science and Engineering Division. The research was performed at Ames Laboratory, which is operated for the U.S. DOE by Iowa State University under contract \# DE-AC02-07CH11358. In addition, W. M. was supported by the Gordon and Betty Moore Foundations EPiQS Initiative through Grant GBMF4411.

\section*{Methods}

\subsection*{Sensor preparation}

An electronic grade single crystalline diamond plate with $\left[100\right]$ surface from Element Six was further thinned down and polished to 40~$\mu$m thickness by {\it Delaware Diamond Knives, Inc.} It was then subjected 14~keV Nitrogen ion irradiation with $10^{15}$~cm$^{-2}$ ion dose by {\it Leonard Kroko, Inc.} According to SRIM calculations, this leads to $\approx$20~nm  projected range of Nitrogen ions diamond with a straggling of 6.5~nm. Nitrogen implanted diamond plates were then subjected to 20~keV energy electron irradiation in an SEM. Consequently, the diamond plates were annealed under $800~^0$C in vacuum for two hours. This mobilizes the vacancies and forms NV centers.\cite{Waldermann07,Toyli10,Schwartz12} Finally, the diamond sensor was subjected to several steps of cleaning including solvent cleaning, acid cleaning with HNO$_3$+HCl 1:3 mixture, and Oxygen plasma cleaning.

\subsection*{Experimental setup}

The experimental setup is based on low temperature atomic force microscope combined with a confocal fluorescence microscope ({\it Attocube} AFM/CFM) in a $^4$He bath cryostat with a base temperature of 4.2~K. Higher temperatures are achieved via a resistive heater and a temperature controller ({\it LakeShore} 335). Optical filters in confocal microscope optimize the NV detection. A low-temperature compatible dry microscope objective ({\it Attocube} LT-ASWDO 0.82 NA) is used in this confocal setup for NV excitation and collection of fluorescence emission. Phonon-mediated fluorescent emission (600-750 nm) of NV centers are detected under coherent optical excitation ({\it Laserglow} R531001FX 532nm LASER) using a single photon counting module ({\it Excelitas} SPCM-AQRH-14). The waist size (diameter) of the excitation laser focus spot is approximately 500~nm. Background static vertical magnetic field is provided by a NbTi superconducting magnet ({\it Cryomagnetics} 4G-100). Microwave (MW) field is generated by a MW synthesizer ({\it Rohde$\&$Schwarz} SMIQ03B), amplified by a 16~W amplifier ({\it Minicircuits} ZHL-16W-43+) and delivered to the NV sensor via a loop antenna formed by a 50~$\mu$m diameter silver wire held between the diamond sample and the microscope objective. A {\it National Instruments} DAQ card (NI PCIe 6323) is utilized for data acquisition.

\section*{Appendix: Decoding the ODMR splittings}

The diamond lattice consists of two interpenetrating face centered cubic Bravais lattices, displaced along the body diagonal of the cubic cell by 1/4 of length of the diagonal. It can be regarded as a face centered cubic lattice with the two-point basis: $(0,0,0)$ and $(\frac{1}{4},\frac{1}{4},\frac{1}{4})$. Here, we assume unit lattice constant for simplicity. The nitrogen-vacancy (NV) center is a point defect in the diamond lattice which consists of a nearest-neighbor pair of a substitution nitrogen atom, and a lattice vacancy.

The four nearest neighbors centered around the lattice point $V_0 = (\frac{1}{4},\frac{1}{4},\frac{1}{4})$ are: $V_1 = (0,0,0), V_2 = (\frac{1}{2},\frac{1}{2},0) , V_3 = (\frac{1}{2},0,\frac{1}{2})$, and $V_4 = (0, \frac{1}{2},\frac{1}{2})$. The four possible NV orientations can therefore be calculated as:
\
\begin{equation}
\hat{d}_i =\frac{V_i - V_0}{|V_i - V_0|} \hspace{1cm}  i=1,..,4
\end{equation}

\begin{equation}
\begin{array} {lcl}
\hat{d}_1 & = & (-1, -1, -1)/\sqrt{3} \\
\hat{d}_2 & = & (1, 1, -1)/\sqrt{3}\\
\hat{d}_3 & = & (1, -1, 1)/\sqrt{3}\\
\hat{d}_4 & = & (-1, 1, 1)/\sqrt{3}\\
\end{array}
\end{equation}

The Zeeman splitting of the $S_z = \pm 1$ states of an NV is given by $2\gamma_e |\vec{B}.\vec{S}|$, where $\gamma_e \approx 2.8~MHz/Oe$ is the gyromagnetic ratio of the NV electronic spin. It is only the magnetic field component that is along the NV orientation lead to Zeeman splitting. Therefore, the possible splittings in an NV ensemble of a single crystalline diamond is given by $2\gamma_e |\vec{B}\cdot\hat{d}|$.

\subsection*{Case I}

For a single crystalline diamond plate with [100] surface placed normal to the field $\vec{B}=(0,0,B_z)$, all the possible NV orientations result in the same splitting:
\begin{equation}
Z = \frac{2 \gamma_e B_z}{\sqrt{3}} ~ \approx ~ 3.233~MHz/Oe
\end{equation}

\subsection*{Case II}

If the magnetic field has two components such that $\vec{B}=(B_x,0,B_z)$, the NV ensemble will result in two pairs of Zeeman splittings. In this case,
\begin{equation}
\begin{array} {lcl}
Z_{L,S} & = & Z|B_z \pm B_x| \\
\\
Max[B_{x,z}] & = & \frac{Z_L + Z_S}{2Z}\\
\\
Min[B_{x,z}] & = & \frac{Z_L - Z_S}{2Z}\\
\end{array}
\end{equation}
Here, $Z_L$ ($Z_S$) refers to larger (smaller) Zeeman splitting.

%\bibliography{NVbib}
%

\end{document}